\documentclass[12pt]{article}
\usepackage{fullpage,setspace}
\usepackage{amssymb, amsthm, amsmath}
\usepackage{graphicx}
\usepackage[authoryear]{natbib}
\usepackage{bm}
\usepackage{adjustbox}
\usepackage{subfig}
\usepackage{graphicx}
\usepackage{float}
\usepackage{booktabs}
\usepackage[format=plain,font=footnotesize,labelfont=bf]{caption}
\usepackage{lineno}
\usepackage{subcaption}
\usepackage{times}
\usepackage{siunitx}
\usepackage{soul}
\usepackage{comment}
\usepackage{xurl}
\usepackage{algorithm2e}
\RestyleAlgo{ruled}

\newcommand{\btheta}{ \mbox{\boldmath $\theta$}}

\newcommand{\bbeta}{ \mbox{\boldmath $\beta$}}

\newcommand{\bX}{ \mbox{\bf X}}

\newcommand{\bZ}{ \mbox{\bf Z}}

\newcommand{\bs}{ \mbox{\bf s}}
\newcommand{\bb}{ \mbox{\bf b}}

\newcommand{\bu}{ \mbox{\bf u}}

\newcommand{\bD}{ \mbox{\bf D}}

\newcommand{\bR}{ \mbox{\bf R}}

\newcommand{\iid}{\stackrel{iid}{\sim}}
\newcommand{\indep}{\stackrel{indep}{\sim}}

\newcommand{\calN}{{\cal N}}

\newcommand{\Matern}{ \mbox{Mat$\acute{\mbox{e}}$rn}}

\newcommand{\beq}{ \begin{equation}}
\newcommand{\eeq}{ \end{equation}}
\newcommand{\beqn}{ \begin{eqnarray}}
\newcommand{\eeqn}{ \end{eqnarray}}

\usepackage{caption}
\renewcommand{\arraystretch}{1.5}

\begin{document}  

\begin{center}
{\Large Spatiotemporal double machine learning to estimate the impact of Cambodian land concessions on deforestation}\\\vspace{6pt}
{\large Anika Arifin\footnote[1]{Smith College}, Duncan DeProfio\footnote[2]{Williams College}, Layla Lammers\footnote[3]{Rhodes College}, Benjamin Shapiro\footnote[4]{University of Florida}, Brian J Reich\footnote[5]{North Carolina State University}, Henry Uddyback$^5$ and Joshua M Gray$^5$}\\
\today
\end{center}

\begin{abstract}\begin{singlespace}\noindent
Environmental policy evaluation frequently requires thoughtful consideration of space and time in causal inference. We use novel statistical methods to analyze the causal effect of land concessions on deforestation rates in Cambodia. Standard approaches, such as difference-in-differences regression, effectively address spatiotemporally-correlated treatments under some conditions, but they are limited in their ability to account for unobserved confounders affecting both treatment and outcome. Double Spatial Regression (DSR) is an approach that uses double machine learning to address these scenarios. DSR resolves the confounding variables for both treatment and outcome, comparing the residuals to estimate treatment effectiveness. We improve upon DSR by considering time in our analysis of policy interventions with spatial effects. We conduct a large-scale simulation study using Bayesian Additive Regression Trees (BART) with spatial embeddings and find that, under certain conditions, our DSR model outperforms standard approaches for addressing unobserved spatial confounding. We then apply our method to evaluate the policy impacts of land concessions on deforestation in Cambodia. \vspace{12pt} \end{singlespace}\end{abstract}

\section{Introduction}\label{s:intro}

Large-scale evaluation of environmental policy frequently relies on observational data with spatial and temporal dependence.  As with all observational studies, drawing casual concludions requires carefully considering confounding variables. 
Causal inference for spatial applications is challenging because spatial dependence violates the common assumptions of causal inference \citep{rubin1974estimating}.  However, novel methodology has recently been developed \cite[for recent reviews, see][]{keller2020selecting, reichReviewSpatialCausal2021, bolin2024infill, dupont2023demystifying}, such as case-control matching, neighborhood adjustments by spatial smoothing, and propensity score weighting.  Most related to our approach is
\cite{wiecha2024two}, who build on the double/debiased machine learning (DML) approach of \cite{chernozhukov2018double} to include a spatial effect on the confounding variable of interest. This approach first removes spatial trends from the outcome and treatment variables and then performs regression on the residuals to estimate the causal effect.  We extend this approach from the spatial to spatiotemporal setting.

Many previous advances in the field of spatiotemporal causal inference have centered around a linear difference-in-differences (DID) model to account for treatment effects across time. However, a standard DID model does not effectively predict a treatment effect when there are correlations in treatment and outcome across nearby locations. Newer methods, however, allow the DID model to account for correlation in treatment across adjacent spaces \citep{delgado2015difference}. This permits the use of DID despite violation of the stable unit treatment value assumption traditionally required in DID, a phenomenon known as “interference.” Even so, spatial causal inference problems often must capture larger-range spatial correlation beyond correlation of adjacent locations. \cite{bardaka2019spatial} achieve progress toward this objective in their spatiotemporal policy evaluation of expanded light rail in Denver, Colorado. They use \cite{delgado2015difference}’s spatial DID model but focus especially on the presence of “spillover effects,” when an intervention in one specific region may have similar effects on nearby areas that it does not directly target. Even after these model innovations, several persistent shortcomings remained in the DID model. Further advances have accounted for effects of spillovers that are not based only on proximity but rely instead upon some other, non-distance factor \citep{butts2021difference}. \citep{butts2021difference} also develops a procedure to adjust for spatial effects of the treatments when the interventions do not occur simultaneously. All of the above methods involve a ``parallel trends" assumption, assuming that treated areas, if not exposed to intervention, would experience the same outcome variable change as untreated areas. \citep{AbadieDID} develop a flexible DID approach can predict treatment effects accurately even when the parallel trends assumption does not hold. This method also does not require observations for the same individuals both before and after treatment. This permits better control of covariates' effects when contrasted with standard DID approaches. Finally, DID has been extended and combined with nonlinear models, seeking to account for confounding variables that change with time \citep{Bhuyan2021}. This methodology utilizes inverse propensity weighted DID estimators for average treatment effects to account for unknown interactions between time and covariates.

Despite advances in DID methods, various biases and shortcomings necessitate the development of other types of analysis to conduct effective spatiotemporal causal inference. One example is the use of a Bayesian Multi-Stage Spatiotemporal Evolution Hierarchy Model (BMSSTEHM) to estimate population-weighted ground-level $O_3$ concentrations in China \citep{Junming-Causal-Inference}. This model combines a Bayesian Spatiotemporal Hierarchal Model (BSTHM) with piecewise regression to model different linear trends before and after the trends shift direction. DML exists as another popular causal inference method. Past simulations have shown that hyperparameter choices can effect the efficacy of DML predictions; for example, crossfitting on the full sample or on multiple folds performs better in small samples than using a split sample approach \citep{Bach2024}. As another alternative to DID, doubly-robust estimators, such as targeted maximum likelihood estimation (TMLE), frequently prove effective in causal inference problems. \cite{Kuhne2022} emphasize these methods and the use of a correct g-formula model to prevent bias. This helps to confront challenges of using causal inference for medical research and health decisions. Also studying fields of health, \cite{Vandenbroucke2016} illustrate potential pitfalls of the Restricted Potential Outcome Approach (RPOA) to causal inference. They suggest that RPOA ranks evidence in a way that neglects important experimental context, resulting in errors in the field of epidemiology. Finally, dimension reduction can provide additional insight into causal inference problems. In climate science, dimension reduction promotes accurate estimation by converting a highly dimensional climate field into a smaller set of regionally constrained patterns for analysis \citep{Falasca2024}.

In the presence of large amounts of unobserved spatial confounding, the DID approach, even with the adjustments described above, often outputs a biased or otherwise inaccurate estimate of the treatment effect. Out of all alternative methods, we select DML for our analysis to improve upon the available DID inference methodologies \citep{chernozhukov2018double}.

Past applications of DML to spatial statistics provide a starting point for our methodology in this spatiotemporal causal inference problem. Our paper builds on \cite{wiecha2024two} to describe the effects of a policy intervention---public land conceded to private businesses---on deforestation in Cambodia. Double Spatial Regression (DSR) is a DML approach that can account for spatially correlated data, but its use has been restricted to analysis at one specific point in time \citep{wiecha2024two}. We apply the methods of DSR to analyze the effect of policy interventions over time accounting for spatial confounding variables in the pre-treatment response, treatment allocation, and post-treatment response. We further expand on \cite{wiecha2024two}, who use Gaussian process regression with linear covariate effects, by using Bayesian Additive Regression Trees (BART) as our machine learning method \citep{BART}. This allows for non-linear covariate effects and fast computation for large spatial datasets. To allow for fidelity to complex spatial trends, we include a spatial embedding layer similar to \cite{chen2024deepkriging}, who add a spatial embedding to a neural network regression. Furthermore, we investigate the consequences of allocating the policy to regions (i.e., polygons) rather than individual locations.  In a simulation study, we compare our spatiotemporal DSR approach to DID, demonstrating increased precision and unbiasedness in accounting for outcomes of policy interventions. We conduct causal inference on the concessions/deforestation relationship, and we conclude with a discussion of future problems and innovations in the field of spatial statistics.

\section{Cambodian deforestation background and data}\label{s:cambodiadata}

Southeast Asian countries like Cambodia have experienced an expansion of rubber farming since the year 2000, coinciding with a rise in prices of the valuable agricultural commodity. The Cambodian government encouraged the development of large plantations during this rubber boom through granting large land concessions to private businesses, many of which replaced natural forests \citep{deforestconsummary}. Nevertheless, deforestation may also occur due to the prevalence of illegal logging in Cambodia and the challenges of enforcing forest protection legislation. For example, in 1997, the illegal logging harvest was over ten times the size of the legal logging output \citep{illegallogging}. This leaves open the question of whether land concessions increase deforestation or merely shift it from illegal loggers to legal companies. Additionally, different crop types may inform amounts of deforestation that are present. For example, rubber plantations often require clear-cutting native forests before planting, whereas timber production frequently reduces deforestation by managing and selectively cutting the existing forest. Other crops grown in Cambodia include black pepper, cashews, and sugar, which can affect rates of deforestation and the regions in which it occurs \citep{deforestconsummary,windsofchange2013}. A number of private companies have emerged to implement and account for these REDD+ projects, but recent analyses suggest that the majority have dubious effectiveness \citep{guizar2022global, west2023action}.

Different strategies exist to combat the loss of biodiversity and ecosystem destruction resulting from deforestation in Cambodia. Protected areas (PAs) maintain biodiversity by sealing off an area from human intervention. \cite{black2022counterfactual} present a study measuring the effectiveness of PAs in preventing deforestation in Cambodia. They use nearest-neighbor propensity scores for their PA assessment. Eight covariates are used to compare treated and untreated samples of data. \cite{black2022counterfactual} find that Cambodia's PAs had a significant, negative association with deforestation in all studied time periods. Forested land in PAs can be up to $12.5 \%$ less likely to face deforestation than matched unprotected forests, though this effect may decrease over time due to increased land concession pressure for agricultural demands. Another method for deforestation reductions is the development of Reduce Emissions from Deforestation and forest Degradation (REDD+) projects. \cite{pauly2022high} study the deforestation differences between REDD+ projects and nearby PAs and find that REDD+ projects were $158\%$ more effective at preventing forest loss than nearby PAs.  

\subsection{Study Area}
\begin{figure}
    \centering
\includegraphics[width=.32\linewidth]{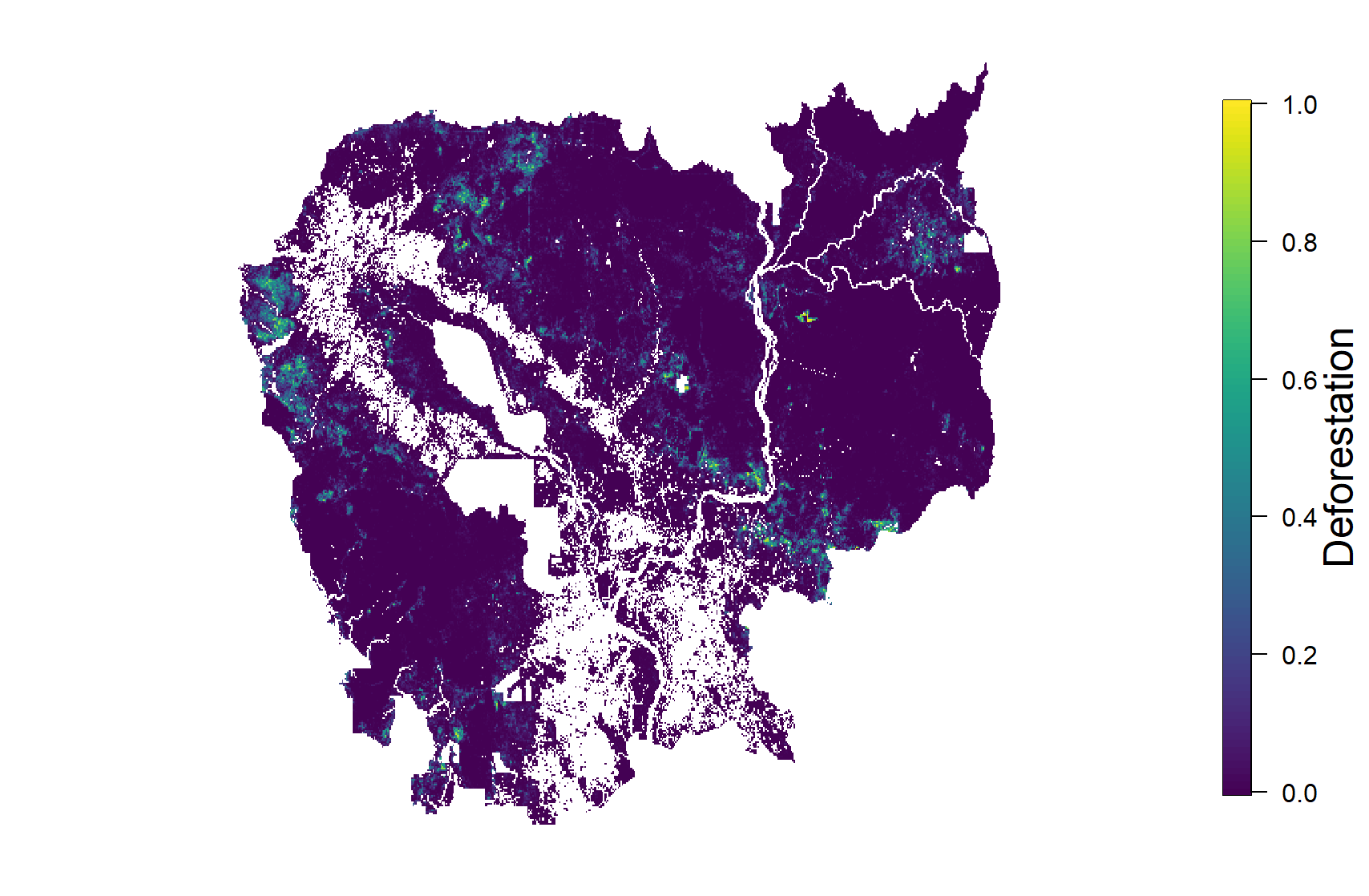}
\includegraphics[width=.32\linewidth]{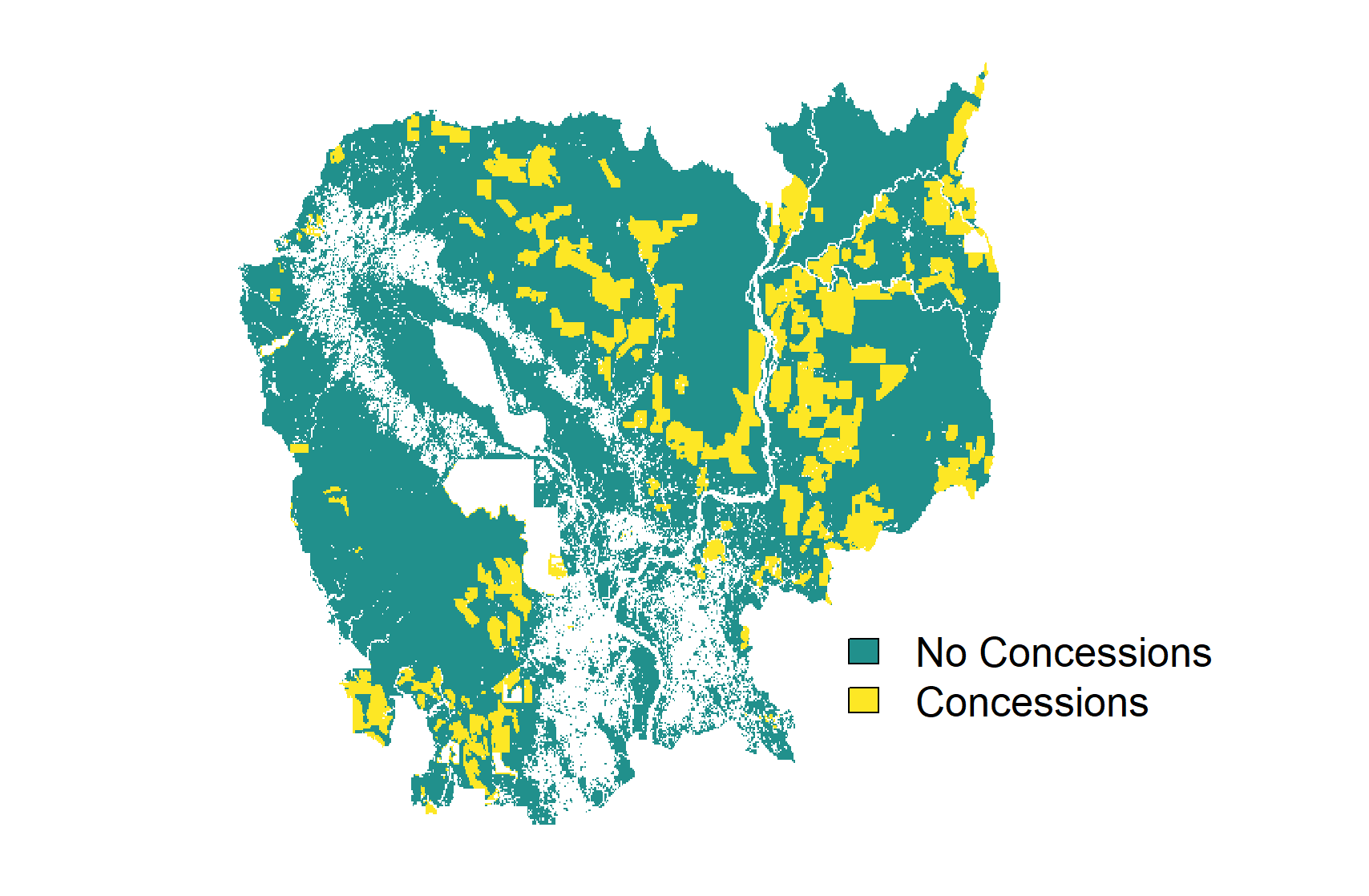} \includegraphics[width=.32\linewidth]{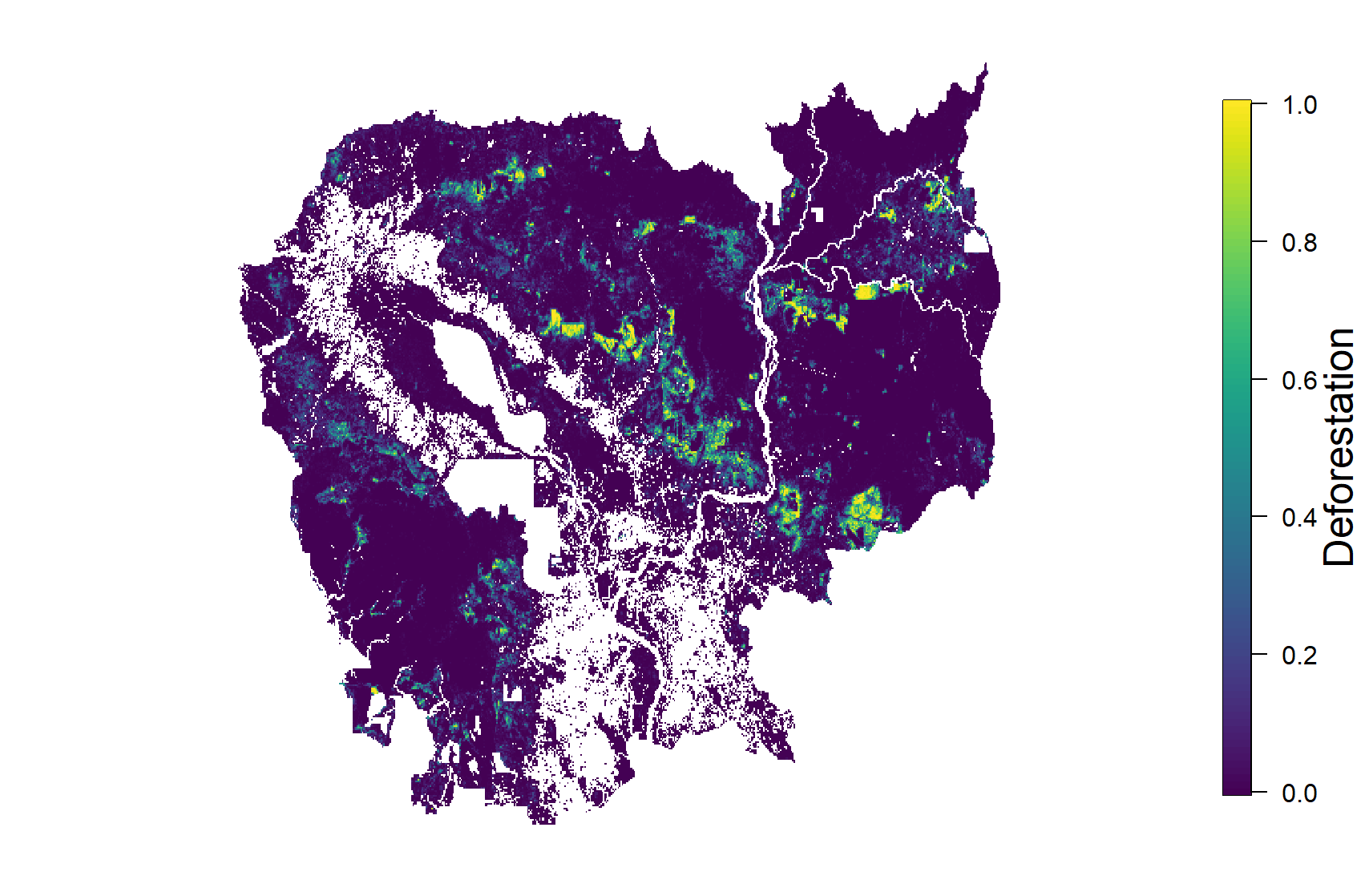}    
   
    \caption{{\bf Deforestation and concessions}: Deforestation by pixel in the pre-intervention (left) and post-intervention (right) period. Purple represents little deforestation, while brighter greens show significant deforestation. A yellow result represents 100 percent deforestation in the raster. A concessions graph is in the center, where yellow indicates the conceded land.}
    \label{fig:exploratory_plots}
\end{figure}

We defined a common 1 km grid covering Cambodia (476 x 580; Figure \ref{fig:exploratory_plots}) and took grid squares as study units. Only grid squares within the Cambodian boundary and having more than 5\% forest cover and less than 5\% urban cover in the year 2000 were retained for analysis (n=145,832). Existing datasets of deforestation, concession/protected status, and covariates (e.g., elevation, population, etc.) were resampled to the common 1 km grid, with an aggregation method appropriate to their use case (details below). Except where noted, data were obtained in their native spatial resolution from Google Earth Engine \citep{GEE} with GDAL \citep{contributors2020geospatial} subsequently used for resampling.

\subsection{Deforestation, Concession and Ancillary Data}
Deforestation data were obtained from the Global Forest Change product \citep[GFC;][]{hansenGFC}, which provides the year of deforestation at 30 m resolution over the entire globe for the period 2000-2024. Annual deforestation area totals were created by aggregating the 30 m deforested pixels to the 1 km grid, in each year, via summation. We also used the GFC's forest coverage in 2000 data to establish a baseline of forest cover in each study unit. 

The Cambodian League for the Promotion and Defense of Human Rights \citep{Con_source} provided the boundaries of forest concessions as well as their type and year of establishment among other (unused) characteristics. Protected area boundaries and types were sourced from the World Database of Protected Areas \citep{UNEPWCMC}. These vector datasets were first rasterized to a 100 m grid using a majority rule for protected/concessed status (binary), year of protected area/concession establishment, and the concession crop type. These datasets were then resampled to the 1 km study grid by averaging the binary protected/concessed status creating a measure of the fraction of the 1 km grid cell that is protected/concessed. The years of establishment and crop type grids were aggregated to the 1 km study grid using the majority rule.

The pre-intervention outcome variable for pixel $i$, $Y_{0}$, is the total forest loss for a pixel during the 4 years prior to, but not including, the concession year. We only assign values of $Y_{0}$ to rasters with a concession date between 2005 and 2024. Likewise, the post-intervention outcome, $Y_{1}$, is the total forest loss spanning the concession year and the three following years. This variable only applies to concessions occurring between 2001 and 2020. Restricted ranges for $Y_{0}$ and $Y_{1}$ result from having available deforestation data only during the years 2001--2023. Figure \ref{fig:exploratory_plots} shows a map of Cambodia with $Y_{0}$ (left) and $Y_{1}$ (right). The pre-treatment mean across pixels for with and without treatment are 0.037 and 0.024, respectively. The corresponding post-treatment means are 0.162 and 0.029. Therefore, a naive difference in differences is (0.162 - 0.029) - (0.037 - 0.024) = 0.12, i.e., a 12$\%$ deforestation of the pixel due to concessions.

The prevalance of deforestation is often dependent on the environmental and human context of a particular place. An area that is far from significant population density, with difficult rugged terrain, and/or with challenging weather conditions may be under less deforestation pressure compared to more easily accessed areas. We summarized this environmental context by creating population, elevation, and climate summaries for each study unit. Population density was obtained from WorldPop and summed within the 1 km study units \citep{Popdata}. The Copernicus global 30 m DEM \citep{elev} was used to compute the average elevation within each study unit. ERA5-Land monthly data \citep{precipTemp} were used to compute the average monthly precipitation and air temperature within each study unit. Additionally, the Copernicus Global Land Cover Layers \citep[CGLS-LC100;][]{urbanwater} were used to sum the areas of urban and water within each grid square.

\section{Double machine learning for spatial difference-in-differences}\label{s:DML}

Let $Y_{it}$ be the observation at location $\bs_i$ for $i\in\{1,...,n\}$ at time $t\in\{0,1\}$ (as described in Section \ref{s:cambodiadata}, some observations are missing), and $\bX_{it} = (1,X_{it1},...,X_{itp})$ be an associated vector of observed covariates (e.g., temperature, precipitation). It is assumed that an intervention is applied to a subset of the locations between the two observation times, with $D_{i}=1$ if site $i$ was given the intervention and $D_i=0$ otherwise.  DML \citep{chernozhukov2018double} is a two-stage estimator that first removes trends in the outcome ($Y_{it}$) and treatment ($D_i$) variables that can be explained by the covariates ($\bX_{it})$ and then regresses the residual outcome onto the residual treatment to estimate the casual effect.  In this section, we extend the spatial DML of \cite{wiecha2024two} to the spatiotemporal DID setting.  The procedure is summarized in Algorithm \ref{alg:dsr}.

\begin{algorithm}
\caption{Spatiotemporal Double Machine Learning}\label{alg:dsr}

\KwData{Outcomes $Y_0$ and $Y_1$, treatment D, spatial location $s$, covariate matrix X, with or without Wendland Basis Expansion}
\KwResult{$\gamma$ estimate}

1. First-stage regressions with cross-fitting\\

Split the locations into $K$ folds

\For {$k = 1,...,K$}{
Create training indices for $K-1$ folds\\
Create testing index for $k^{th}$ fold\\
Fit $Y_0$ and $Y_1$ separately using {\tt bart} from {\tt dbarts} \citep{dbarts}\\
Predict $\hat{Y_0}$ and $\hat{Y_1}$  for the $k^{th}$ fold\\
Fit $D$ using ${\tt pbart}$ from ${\tt BART}$ \citep{BART}.\\
Predict $\hat{D}$ for the $k^{th}$ fold}

2. Second-stage regression\\
Compute the residuals\\
\quad$R_{i0}^Y = Y_{i0} -\hat{Y_{i0}}$, 
\quad$R_{i1}^Y = Y_{i1} -\hat{Y_{i1}}$, \mbox{\ \ and \ \ }
\quad$R_{i1}^D = D_{i} -\hat{D_{i}}$\\
Use the residuals to estimate $\gamma$ in second-stage regression\\
\quad$R_{it}^Y = \beta + \delta t + \alpha R_i^D + {\bar \alpha} {\bar R}_i^D + \gamma tR_i^D + {\bar \gamma} t{\bar R}_i^D + \varepsilon_{it}$
\end{algorithm}

\subsection{Stage 1: Spatial regressions}\label{ss:DML1}

The first stage builds spatial regression models for the outcome and treatment variables.  The outcome model for time $t\in\{0,1\}$ is 
\begin{equation}\label{e:first_stageY}
 Y_{it} = f_{t}(\bX_{it},\bs_i) + \epsilon_{it}
\end{equation}
where $f_t$ is an unknown mean function, $\epsilon_{it}\indep\mbox{Normal}(0,\sigma_t^2)$ is error and the regressions are fit separately for each time point.  The binary treatment variable is modeled as the spatial logistic regression model with pre-treatment covariates $\bX_{i0}$,
\begin{equation}\label{e:first_stageD}
 \mbox{logit}\{\mbox{Prob}(D_{i} =1)\} =g(\bX_{i0},\bs_i).
\end{equation}
for unknown function $g$.

The functions $f_0$, $f_1$ and $g$ can be estimated with any machine learning method, including Gaussian process regression, i.e. Kriging, or neural networks.  In this paper, we use Bayesian Adaptive Regression Trees (BART; \cite{BART}) to provide arbitrarily flexible models for the unknown regression functions that can be fit with large datasets with continuous $(Y_{it})$ and binary ($D_i$) outcomes, as well as random effects (see Section \ref{ss:DMLre}). BART builds a model of data by developing a set of binary decision trees based on its input variables. This machine learning method uses the end points of these trees to develop a posterior distribution that models the data.

Including the coordinates $\bs_i$ as features in the BART regression can capture complex spatial trends.  However, for large and diverse regions such as Cambodia, this may require a large number of deep trees.  To facilitate spatial modeling, we follow \cite{chen2024deepkriging} and use an embedding layer to supplement the coordinates with a spatial basis expansion.   Following \cite{chen2024deepkriging}, we use define a rectangular grid of knots location $\bu_1,...,\bu_L$ covering the spatial domain of interest and use Wendland basis functions \citep{nychka2015multiresolution}
\begin{equation}\label{e:wendland}
Z_{il} = \begin{cases}
(1-d_{il})^6(36d_{il}^2 +18d_{il}+3)/3 & \text{if } d_{il}\le 1\\
0 & \text{if } d_{il}>1
\end{cases}\end{equation}
for $d_{il} = ||\bs_i - \bu_l||/\phi$ and bandwidth parameter $\phi$ set to 2.5 times the grid spacing. The basis functions $\bZ_i = (Z_{i1},...,Z_{iL})^T$ are then used as inputs to BART.  That is, we fit the model with means $f_{t}(\bX_{it},\bs_i, \bZ_i)$ and $g(\bX_{i0},\bs_i,\bZ_i)$.

If the functions $f_t$ and $g$ are smooth in space, the first-stage regression can be fit to the entire training data \citep{wiecha2024two}. Otherwise, cross-fitting is a way to ensure the first-stage regressions do not overfit and obscure the intervention effect.  In cross-fitting, the $n$ locations are randomly partitioned into $K$ folds, and the predictions ${\widehat Y}_{i0}$, ${\widehat Y}_{i1}$ and ${\widehat D}_i$ are made from a model trained on the data from the $K-1$ folds that exclude site $i$. In the analyses below we take $K=10$ and allocate sites (the same allocation for all three regressions) into folds using a complete random sample.  

\subsection{Stage 2: Final estimate and standard error}\label{ss:DML2}

Denoting the cross-fitting residuals as $R_{i0}^Y = Y_{i0} - {\widehat Y}_{i0}$, $R_{i1}^Y = Y_{i1} - {\widehat Y}_{i1}$ and $R_{i}^D = D_{i} - {\widehat D}_{i}$, the second-stage regression is  \begin{equation}\label{e:DMLregression}
   R_{it}^Y = \beta + \delta t + \alpha 
   R_i^D + {\bar \alpha} {\bar R}_i^D + \gamma tR_i^D + {\bar \gamma}
   t{\bar R}_i^D + \varepsilon_{it}, 
\end{equation}
where ${\bar R}_i^D$ is the local mean of $R_i^D$ over the neighboring pixels.  In (\ref{e:DMLregression}), the covariates $\bX_i$ are removed because their effects have been removed in the first stage.  The treatment effect of interest, $\gamma$, is obtained by least squares regression. Letting $\bR$ be the vector length $2n$ composed of the $R_{it}^Y$, $\bZ$ be the corresponding $2n\times 6$ design matrix and $\btheta = (\beta,\delta, \alpha, {\bar \alpha},\gamma, {\bar \gamma})^T$ be the unknown coefficients, the estimator is 
$${\hat \btheta} = (\bZ^T\bZ)\bZ^T\bR.$$

As in \cite{chernozhukov2018double} and \cite{wiecha2024two}, the standard error is computed using the robust method of \cite{mackinnon1985some},
$$\mbox{Cov}({\hat \btheta}) = (\bZ^T\bZ)^{-1}\bZ^T\bD\bZ(\bZ^T\bZ)^{-1}$$
where $\bD$ is the diagonal matrix with elements 
$(R_{it}^Y - {\hat \beta} - {\hat \delta} t - {\hat \alpha} 
   R_i^D - {\hat {\bar \alpha}} {\bar R}_i^D - {\hat \gamma} tR_i^D - {\hat {\bar \gamma}}
   t{\bar R}_i^D)^2$.  This is computed in the {\tt lm$\_$robust} function in {\tt R} with the ``HC0'' standard error option \citep{estimatr}.  A heuristic mathematical argument that this two-stage algorithm can remove spatially-smooth missing confounder variables is given in Appendix A. Code to implement the method is available at \url{https://github.com/ddeprof/Spatiotemporal_DML}.

\subsection{Extension to clustered treatments}\label{ss:DMLre}

In concessions within Cambodia, the government grants contiguous polygons to private entities. Hence, treatments can be said to be ``clustered" into discrete groups. We can group treated pixels into polygons where they undergo the same treatment status as a unit. We can model each polygon as a block with random effects (RE) on the outcome variable $Y$. This reflects the reality that each piece of land may have different qualities and circumstances that affect the deforestation outcome. For example, some companies may eagerly deforest all land, clear-cutting the entire concession with great precision, while others may have incentives to preserve some or all of the forested land on their new property. This modeling of clustered treatments is further explored in a detailed simulation study (see Section \ref{s:sim}) to test the efficacy of ST-DML methodology on more realistic data scenarios.

\section{Simulation study}\label{s:sim}

We conduct two simulation studies to evaluate the performance of the proposed DML methods.  In Section \ref{s:sim:pixel}, treatment is assigned at the pixel level.  To mimic the Cambodia data, in Section \ref{s:sim:block} treatment is assigned to blocks of pixels.  In both cases, we generate data on an $m\times m$ grid of spatial locations spanning $[0,1]^2$.  The data-generating model includes five spatial covariates.  For covariate $j$, $(X_{1j},...,X_{nj})$ is generated as a Gaussian process with mean zero, variance one, and $\Matern$ correlation with range $\rho$ and smoothness $\nu$ using the package {\tt spectralGP} \citep{spectralGP}.  The covariates are independent over $j\in\{1,...,5\}$ and shared by the treatment and outcome models.  In the analysis of the simulated data, it is assumed that the first $p=3$ covariates are observed and included in analysis as $\bX_{i} = (1,X_{i1},...,X_{ip})$, but the remaining covariates are not observed.  Since these missing covariates are associated with both the treatment and response, this omission induces spatial confounding.

\subsection{Simulation design with pixel-level treatment allocation}\label{s:sim:pixel}

Given the covariates $\bX_i$, $\mbox{logit}\{\mbox{Prob}(D_i=1)\} = h_1(\bX_i)$ and $Y_{it} = h_2(t,\bX_{i},D_i) + \varepsilon_{it}$ for $t\in\{0,1\}$,  with $\varepsilon_{it} \iid\mbox{Normal}(0,\sigma^2)$.  The regression functions are taken to be \citep{BartPackageR}
\begin{eqnarray}\label{e:trueh}
h_1(\bX_i)&=& \sin(\pi X_{i1}X_{i2}) + 20(X_{i3} - 0.5)^2 + 10X_{i4} + 5X_{i5}\\
\label{e:trueh2} h_2(t,\bX_i,D_i)&=& X_{i1} + tX_{i1} + 3X_{i4} + 5tX_{i}5 +  \gamma t D_i.
\end{eqnarray}
 The interaction between covariates and time in (\ref{e:trueh2}) breaks the parallel trends assumption required for competing methods OLS and DID. The true treatment effect is set to $\gamma=3$.  The other parameters are set to $\rho = 0.3$, $\sigma^2 = 1$ and $m=32$ so $n=m^2=1024$.  A complete random sample of 20\% of the $Y_{it}$ are taken to be missing.  We vary $\nu\in\{1,2,5\}$ to explore how changes in the smoothness affect the efficacy of our ST-DML method.  We simulate 120 datasets for each $\nu$.
 
For each dataset, we fit six versions of the ST-DML method. We vary the feature sets in the first-stage BART regressions: $\bX_i$ only (``X''), $\bX_i$ and $\bs_i$ (``XS''), and $\bX_i$, $\bs_i$ and $\bZ_i$ with $L=100$ basis functions (``XSZ''). Each method is fit with and without crossfitting. For BART on outcome $Y_0$ and $Y_1$, we use the default settings in the {\tt{bart}} function from the {\tt{dbarts}} package \citep{dbarts}; for BART on the treatment $D$, we use the default settings in the {\tt pbart} function from the {\tt BART} package \citep{BART}. We compare ST-DML with the spatiotemporal difference-in-differences (ST-DID) method proposed by  \cite{delgado2015difference}.  They include the mean of the treatment at neighboring locations as a local confounder adjustment.  Let
${\cal N}_i$ be the collection of $m_i$ sites deemed as neighbors of site $i$, and ${\bar D}_{i} = \sum_{j\in\calN_i}D_j/m_i$ be the proportion of the neighbors given the intervention. This local treatment adjustment is added 
 as
\begin{equation}\label{e:nonspatialDID}
   Y_{it} = \bX_{it}\bbeta + \delta t + \alpha D_i + {\bar \alpha} {\bar D}_i + \gamma tD_i + {\bar \gamma}
   t{\bar D}_i + \varepsilon_{it}. 
\end{equation}
We also compare this with an unadjusted ordinary least squares (OLS) fit that sets $\bar{\alpha} = \bar{\gamma} = 0$. For both methods, the interaction effect $\gamma$ is causal effect of interest, estimated using least squares.  Methods are compared with bias, mean squared error (MSE), average confidence interval length and coverage of 95\% confidence intervals for the treatment effect, $\gamma$.


\subsection{Simulation design with clustered treatment allocation}\label{s:sim:block}

In our motivating application, treatment is allocated to spatially continuous blocks.  In this simulation, we test whether our pixel-level spatial model can accommodate this block treatment allocation. We partition the $n$ pixels into $G=64$ blocks, each a $4\times 4$ rectangle.  Let $g_i\in\{1,...,G\}$ be the block that contains pixel $i$.  Treatment is allocated at the block level using the block average of the covariates.  For $j\in\{1,...,p\}$ and $g\in\{1,...,G\}$, let ${\bar X}_{gj}$ be the mean of $X_{ij}$ in block $g$ and ${\bar \bX}_g = (1,{\bar X}_{g1},...,{\bar X}_{gp})$.  Then the treatment in block $g$ is generated as $\mbox{logit}\{\mbox{Prob}({\bar D}_g=1)\}  = h_1({\bar \bX}_g)$ and individual pixels are assigned the block treatment, $D_i = {\bar D}_{g_i}$. The observations are then generated as  \begin{equation}\label{e:BARTre}
  Y_{it} = \alpha_{g_i} + h_{2}(t,\bX_{i},D_i) +\varepsilon_{it}
\end{equation}
where $\alpha_g\iid\mbox{Normal}(0,\tau^2)$ , $\varepsilon_{it} \iid\mbox{Normal} (0, \sigma^2)$, and $\tau^2=\sigma^2=0.25$, where functions $h_1$ and $h_2$ are in (\ref{e:trueh}) and (\ref{e:trueh2}).  There are no missing observations and we simulate 120 datasets.

We test six versions of the ST-DML approach. All methods include $\bX_i$, $\bs_i$ and $L=100$ Wendland basis functions $\bZ_i$ as features in BART.  We consider three methods of crossfitting: no crossfitting and crossfitting ``by pixel" and ``by block." In crossfitting by pixel, we allocate pixels to the $K = 10$ folds, as in the simulation described on data without clustered treatment, but for crossfitting by block we randomly assign the $G$ blocks to the $K$ folds and allocate pixels according to their block's assignment. In the second stage for block crossfitting, we modify equation (\ref{e:DMLregression}) to be $R_{it}^Y = \beta + \delta t + \alpha 
   R_i^D + \gamma tR_i^D + \varepsilon_{it}.$
We perform all 3 crossfitting methods using BART with and without random effects in the first stage. In crossfitting, we integrate random effects by block as $Y_{it}=\alpha_{g_it} + f_t(\bX_{it},\bs_i)+\epsilon_{it}$ with $\alpha_{gt}\indep\mbox{Normal}(0,\tau_t^2)$ separately for $t \in\{0,1\}$ and $\mbox{logit}\{\mbox{Prob}({\bar D}_g=1)\}  = g({\bar \bX}_g)$. This random-effect model is fit using the function {\tt rbart$\_$vi} in the package {\tt dbarts} for BART on the outcome variable \citep{dbarts}, whereas the treatment variable's fit continues using {\tt pbart} from the {\tt BART} package.

\subsection{Results}\label{s:sim:results}

The results of the pixel-level simulation with $\nu=2$ are given in Table \ref{tab:noRE-smooth2}; see Appendix B for the results with the other smoothness parameters.   For all three smoothness levels, competing methods OLS and DID suffer from large bias and MSE and low coverage. The ST-DML methods that use spatial information in BART are more efficient than non-spatial methods, with ST-DML with cross-fitting and Wendland basis expansion in the first stage performing best of all compared methods. With the exception of the methods that exclude spatial information, the bias and MSE are both lower for the ST-DML methods with crossfitting compared to their corresponding methods without crossfitting. From these simulations, we conclude that using BART with the Wendland basis expansion and crossfitting provides our strongest estimation for treatment effect $\gamma$. Although this is true across all smoothness values, our suggested ST-DML method grows more accurate as smoothness increases (see Appendix).

\begin{table} [h!]
    \centering
    \caption{Comparison of ordinary least squares (OLS), difference-in-differences (DID) and spatiotemporal double ML (DML) for {\bf pixel-level treatment allocation and smoothness $\nu=2$}.  DML methods vary by excluding spatial information (X), using spatial coordinates (XS) and spatial basis functions (XSZ) in BART fits and performing DML with and without cross-fitting (CF).  Methods for estimating the treatment effect are compared using bias, mean squared error (MSE), confidence interval (CI) length and coverage of 95\% intervals. } 
    \renewcommand{\arraystretch}{1}
    \begin{tabular}{lcccc}
    \toprule
        \textbf{Method} & \textbf{Bias} & \textbf{MSE} & \textbf{CI length} & \textbf{Coverage}  \\
    \midrule
        OLS & \hspace{5.5pt} $0.821$ & $1.191$ & $1.578$ & $0.433$ \\
        DID  & \hspace{5.5pt} $0.535$ & $0.536$ & $1.839$ & $0.825$  \\
        DML - X - no CF & \hspace{5.5pt} $0.485$ & $0.457$ & $1.687$ & $0.800$ \\
        DML - X - CF & \hspace{5.5pt} $0.643$ & $0.633$ & $1.771$ & $0.733$ \\
        DML - XS - no CF & $-0.614$ & $0.445$ & $0.724$ & $0.133$ \\
        DML - XS - CF & \hspace{5.5pt} $0.223$ & $0.123$ &$1.070$ & $0.883$ \\
        DML -  XSZ - no CF & $-0.763$ & $0.624$ & $0.583$ & $0.017$ \\
        DML - XSZ - CF & \hspace{5.5pt} $0.083$ & $0.044$ & $0.853$ & $0.933$ 
    \end{tabular}
    \label{tab:noRE-smooth2}
\end{table}

We proceed to the simulation with block treatment allocation as described in Section \ref{s:sim:block}.  Table \ref{tab:RE_andno_RE_results} and Figure \ref{fig:RE_andnoRE_comparison} (see Appendix A.1) compare ST-DML methods with OLS and DID.  ST-DML performs more accurately with crossfitting but without random effects integrated in BART. Crossfitting by block yields the lowest bias, but coverage and MSE are better when crossfitting by pixel. These two methods outperform competing methods OLS and DID, as well as other ST-DML methods that do not have crossfitting or that utilize random block effects directly. From these comparisons, we conclude that DML with Wendland basis expansion and crossfitting by pixel, without random effects, is the strongest methodology for data with clustered treatment allocation. 

\begin{table} [h!]
    \centering
    \caption{Comparison of ordinary least squares (OLS), difference-in-differences (DID) and spatiotemporal double ML (DML) for {\bf block-level treatment allocation}.  DML methods are fit with and without block random effects (RE) in the first stage and with no cross-fitting and cross-fitting by pixel and block.  Methods for estimating the treatment effect are compared using bias, mean squared error (MSE), credible interval (CI) length and coverage of 95\% intervals. } 
    \renewcommand{\arraystretch}{1}
    \begin{tabular}{lccccccc}
    \toprule
        \textbf{Method} & \textbf{Bias} &  \textbf{MSE} & \textbf{CI length} & \textbf{Coverage} \\
    \midrule
        OLS & \hspace{5.5pt} 0.664  &  1.686 & 1.614 & 0.525 \\
        DID  & -- 0.384 & 0.881 & 4.558 &0.983  \\
        DML - no RE - no CF & -- 0.844 & 0.824 &1.330  & 0.267  \\
        DML - no RE - pixel CF  & -- 0.197 & 0.112 & 1.413 & 0.983 \\
        DML - no RE - block CF  & \hspace{5.5pt} 0.132 & 0.352  &0.923  &0.583  \\
        DML - RE - no CF & \hspace{5.5pt} 1.133 & 1.585 & 0.981 & 0.133 \\
        DML - RE - pixel CF & \hspace{5.5pt} 1.148 & 1.978 & 2.849 & 0.650 \\
        DML - RE - block CF & \hspace{5.5pt} 0.486& 1.396 & 1.803 & 0.583 \\
    \end{tabular}
    \label{tab:RE_andno_RE_results}
\end{table}


\section{Application to Cambodian deforestation}\label{s:app}


To analyze the Cambodian deforestation data, following the results in Section \ref{s:sim}, we use ST-DML with crossfitting by pixel and without random effects. 
We first analyze the relative variable importance of different confounders that affect treatment and outcome in the first-stage BART fits. The statistic {\tt varcount} from the {\tt dbarts} package measures variable importance during our BART fitting in the first stage of ST-DML ($L$ = 100) and is recorded in Table \ref{tab:Variable_Importance_100}. We average the {\tt varcount} statistic across all $K$ folds. Forest cover is one of the most important confounding variables affecting both treatment and outcome. This is reasonable because there should not be large amounts of deforestation where there are few trees. Likewise, barren lands are often less desirable for farming and logging than fertile, forested ones -- leading to less concessions associated with lands of less forest cover. Climate variables precipitation and temperature also prove important, likely because they affect the growth of different crops. On the other hand, population, urbanization, and water variables are less important in our stage-one modeling. They have less impact on deforestation and concessions.

\begin{table} [h!]
    \centering
    \caption{Variable Importance for the BART fits using $L$ = 100 spatial basis functions} 
    \renewcommand{\arraystretch}{1}
    \begin{tabular}{lccccccc}
    \toprule
        \textbf{Variable} & \textbf{Forest Cover} &  \textbf{Elevation} & \textbf{Population} & \textbf{Precip} & \textbf{Temp} & \textbf{Urban} & \textbf{Water} \\
    \midrule
        $Y_0$ & 66.5 & 18.4 & 6.0 & 15.7 & 13.8 & 7.0 & 3.9 \\
        $Y_1$  & 60.7 & 17.1 & 3.8 & 14.4 & 16.7 & 1.4 & 4.8\\
        $D$ & 19.8 & 11.2 & 3.6 & 9.9 & 11.6 & 1.6 & 1.2 \\
    \end{tabular}
    \label{tab:Variable_Importance_100}
\end{table}

\begin{figure}[h!]
    \centering
    \includegraphics[width=0.6\linewidth]{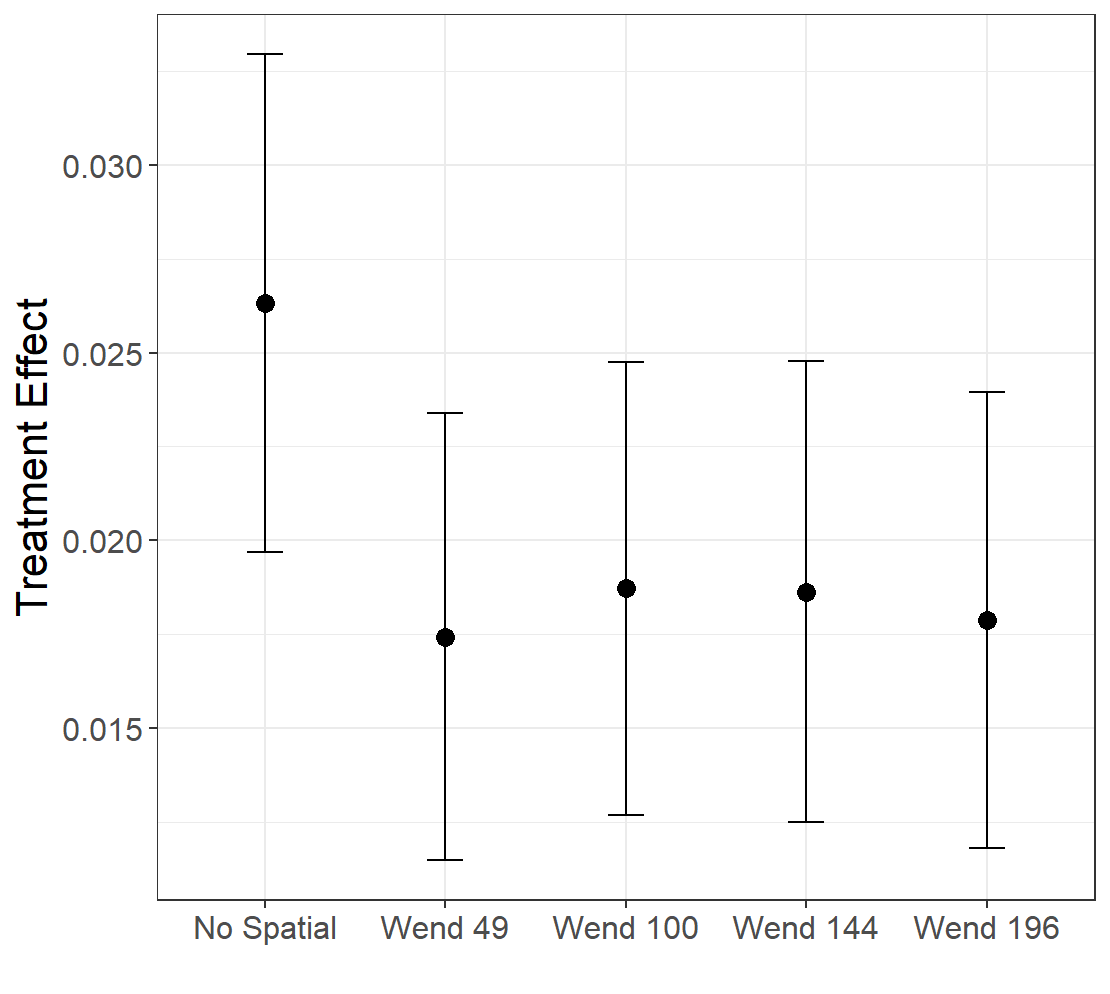}
    \caption{Treatment effect estimation using varying levels of spatial information, from no spatial information to $L$ = 196 Wendland basis functions. The points are estimates, and bars are 95$\%$ confidence intervals.}
    \label{fig:Knot_comparison}
\end{figure}

When we estimate the effect of land concessions on deforestation, we witness a positive treatment effect (i.e., concessions slightly increase deforestation) for all methodologies (Table \ref{tab:DML_Model_Spatial}). However, compared to OLS and DID, we have a much smaller treatment effect from the ST-DML method that more effectively removes spatial confounders. We compare the estimated treatment effect as we vary the number of knot points in the Wendland basis expansion ($L$) for ST-DML's stage one BART input. Figure \ref{fig:Knot_comparison} demonstrates that including the spatial basis functions reduces the treatment effect, but the results are not sensitive to the number of basis functions. According to our full model involving crossfitting and Wendland dimension expansion with $L=100$, a raster conceded in Cambodia experiences about 1.8$\%$ of extra deforestation on average than one not conceded. This is a small number; potential reasons limiting its magnitude include rampant illegal deforestation on non-conceded lands or companies waiting long periods of time to begin deforestation. This demonstrates a small but statistically significant effect of land concessions on deforestation in Cambodia.   

\begin{table} [h!]
    \centering
    \caption{Effect estimates for the Cambodia data using ordinary least squares (OLS), spatiotemporal Differences in Differences (DID) and double machine learning without (DML - nonspatial) and with (DML - spatial) including spatial covariates ($\bs$ and $L=100$ Wendland basis functions) in BART fits.} 
    \renewcommand{\arraystretch}{1}
    \begin{tabular}{lccccccc}
    \toprule
        \textbf{Method} & \textbf{Estimate} &  \textbf{Standard Error} & \textbf{CI Lower} & \textbf{CI Upper} \\
    \midrule
        OLS & 2.63 & 0.34& 1.97 & 3.29 \\
        DID  & 1.74 & 0.30 & 1.15 & 2.34 \\
        DML - nonspatial & 1.87 & 0.31 & 1.27 & 2.48  \\
        DML - spatial  & 1.86 & 0.31 & 1.25 & 2.48 \\
     \end{tabular}
    \label{tab:DML_Model_Spatial}
\end{table}

\section{Discussion}\label{s:discussion}

The proposed spatiotemporal double machine learning approach addresses missing confounders in evaluating policy interventions with spatiotemporal observational data. We propose a flexible model using Bayesian Additive Regression Trees (BART) with a spatial embedding layer to account for missing spatial covariates present before and after treatment.  A simulation study was conducted with two designs to evaluate the performance of the DML methods. The first design is with treatment allocation at the pixel level and the second is with clustered treatment allocation. For both studies, when addressing unobserved spatial confounding, our ST-DML model outperforms the leading approaches. Our application of ST-DML to evaluate the policy impacts of land concession on deforestation in Cambodia finds that for when treatments are observed in Cambodia, the average amount of deforestation increases by 2$\%$.

Limitations in this study design provide enticing areas for future research. For example, we assumed a homogeneous treatment effect, but the effect of land concessions may in fact exhibit differences in various regions of Cambodia (i.e., crop type). We could consider statistics like the conditional average treatment effect that provide insight on potential treatment heterogeneity. Similarly, our simulated model of treatment allocation (based on equally-sized, square blocks) does account for clustered treatments, but it fails to account for the diversity of sizes and shapes of land concessions. Our simulated treatment clusters were defined solely on basis of spatial proximity, while real land polygons may have other confounders in common that influence concession decisions. Future work can include treatment allocation models that more accurately capture real-world concession processes. 

Finally, future work can incorporate further novel perspectives on our inputs of space and time. Currently, we account for time as a binary variable, where we consider ``before" as the 4 years pre-treatment, and ``after" as the concession year and the following 3 years. However, this fails to consider how some companies may take longer than 4 years to begin their processes of deforestation (potentially contributing to our relatively small 2$\%$ treatment effect). This, along with the presence of rampant illegal deforestation on non-conceded lands, could explain how the treatment effect of $1.8\%$ is relatively small in magnitude. Another potential step could be consideration of time in 3 or more different stages or as a continuous variable. Regarding space, we considered neighborhood effects during our analysis, but we did not account for larger-scale spillover effects that still occur outside of adjacent squares. Incorporating this spillover could be an interesting next step in investigating this policy evaluation problem.

\section*{Acknowledgements} 

This work was supported by the National Science Foundation (DMS2349611) and National Institutes of Health (1R01ES036270-01A1).

\begin{singlespace}
	\bibliographystyle{rss}
	\bibliography{refs}
\end{singlespace}

\clearpage

\section*{Appendix A: Removal of spatial confounders}\label{s:B1}

Here we provide a heuristic justification that the proposed estimator can account for spatially-smooth unmeasured confounders. We make the following assumptions:

\noindent\textbf{(A1)} The data-generating model is
\begin{equation*}
    Y_{it} = \theta_{it} + t\,\gamma\, D_i + \epsilon_{it}.
\end{equation*}
\noindent\textbf{(A2)} The spatial processes are decomposed as
\begin{align*}
    \theta_{it} &= u_{it} + e_{it}, \quad t \in \{0,1\}, \\
    D_i &= u_{i2} +e_{i2}
\end{align*}
where $\mathbf{u}$ are spatially smooth and $\mathbf{e}$ is spatially-independent local error.

\noindent\textbf{(A3)} The first stage estimators can fit the spatially smooth components. Thus,
\begin{align*}
    \hat Y_{it} &= u_{it} + \gamma t u_{i2},
    \quad t \in \{0,1\}, \\
    \hat D_i &= u_{i2}.
\end{align*}

\noindent\textbf{(A4)} The local errors satisfy
\begin{equation*}
    \mathbb{E}\left[\sum_{i=1}^{n} e_{i2} \left(e_{i1} + \epsilon_{i1}\right) \,\middle|\, \mathbf{e}_2\right]=0.
\end{equation*}

\noindent BART is fit in the first stage to obtain estimators $\hat Y_{i0}$, $\hat Y_{i1}$ and $\hat D_i$. With the above assumptions, the residuals from the regression are
\begin{align*} 
    R_{i0} &= Y_{i0} - \hat Y_{i0} = \left[u_{i0} + e_{i0} + \epsilon_{i0}\right] - u_{i0} = e_{i0} + \epsilon_{i0}, \\
    R_{i1} &= Y_{i1} - \hat Y_{i1} = \left[u_{i1} + e_{i1} + \gamma \left(u_{i2} + e_{i2} \right) + \epsilon_{i1}\right] - \left[u_{i1} + \gamma u_{i2} \right] = e_{i1} + \epsilon_{i1} + \gamma e_{i2}, \\ 
    R_{i2} &= D_i - \hat D_i = \left[u_{i2} + e_{i2}\right] - u_{i2} = e_{i2}.
\end{align*}

\noindent The ordinary least-squares estimator $ \hat \gamma$ is
\begin{equation}
    \hat{\gamma} = \frac{\sum_{i=1}^{n} R_{i1} R_{i2}}{\sum_{i=1}^{n} R_{i2}^2} 
    = \frac{\sum_{i=1}^{n} \left[e_{i1} + \epsilon_{i1} + \gamma e_{i2} \right] e_{i2}}{\sum_{i=1}^{n} e_{i2}^2} \\
    = \gamma + \frac{\sum_{i=1}^{n} \left[e_{i1} + \epsilon_{i1} \right] e_{i2}}{\sum_{i=1}^{n} e_{i2}^2}.
\end{equation}

\noindent For $\hat{\gamma}$ to be an unbiased estimator,
\begin{align*}
    0 &= \mathbb{E}\left[\frac{\sum_{i=1}^{n} \left(e_{i1} + \epsilon_{i1} \right) e_{i2}}{\sum_{i=1}^{n} e_{i2}^2}\right] \\
    &= \mathbb{E}\left[\mathbb{E}\left[\frac{\sum_{i=1}^{n} \left(e_{i1} + \epsilon_{i1} \right) e_{i2}}{\sum_{i=1}^{n} e_{i2}^2} \,\middle|\, \mathbf{e}_2 = \{e_{12}, e_{22}, \ldots, e_{n2}\}\right]\right] \\
    &= \mathbb{E}\left[\frac{\mathbb{E}\left[\sum_{i=1}^{n} e_{i2}(e_{i1}+\epsilon_{i1}) \,\middle|\, \mathbf{e}_2\right]}{\sum_{i=1}^{n} e_{i2}^2}\right].
\end{align*}

\noindent The denominator $\sum_{i=1}^{n} e_{i2}^2>0$ and the numerator is zero by Assumption A4, so the bias is zero.  Of course, this result hinges on Assumption A3 that spatial trends are exactly removed by the first-stage regression, which is unrealistic, but the argument presented here builds the intuition for the two-stage estimator.

\section*{Appendix B: Additional simulation study results}\label{s:A1}

\begin{figure}[h!]
    \centering
    \includegraphics[width=0.8\linewidth]{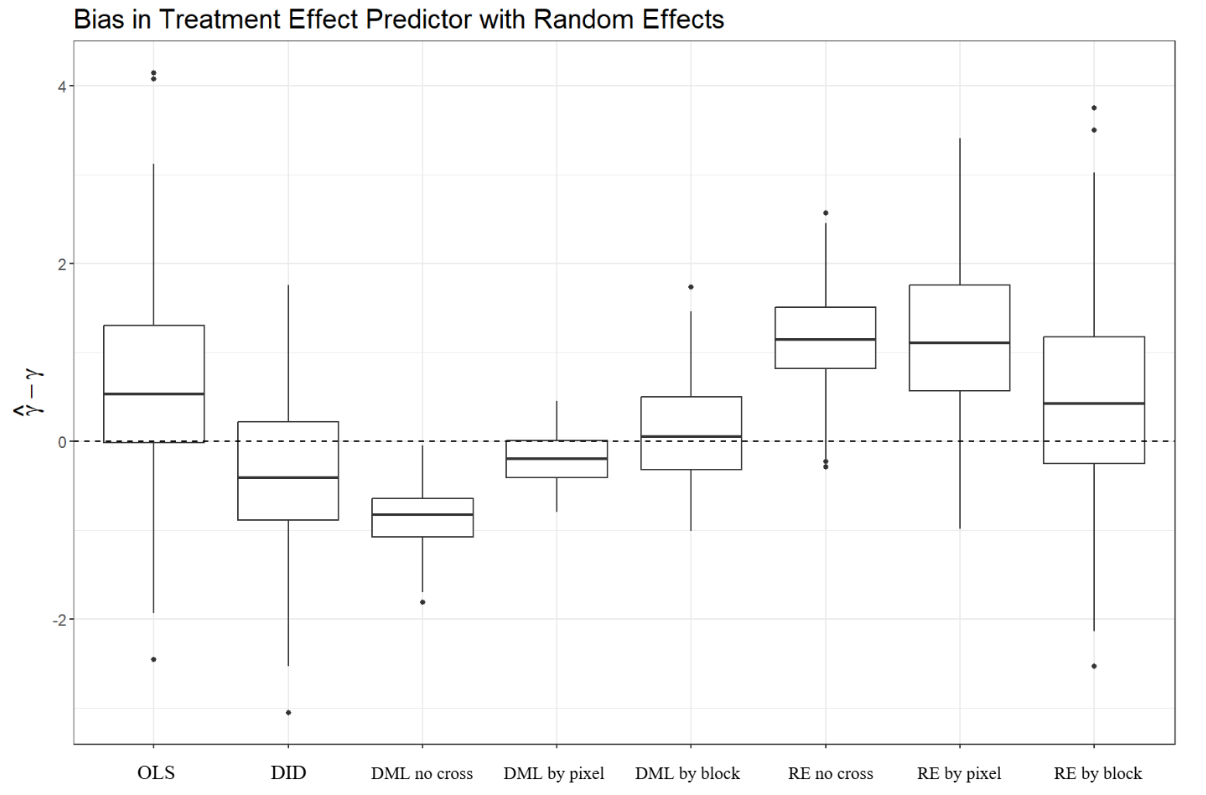}
    \caption{Simulation study comparing OLS, DID, three methods without random effects, and three methods with random effects. The true value of $\gamma=3$, and smoothness $\nu=2$. }
    \label{fig:RE_andnoRE_comparison}
\end{figure}

Tables \ref{tab:noRE-smooth1} and \ref{tab:noRE-smooth5} give the results of the pixel-level treatment allocation simulation with smoothness parameters $\nu=1$ and $\nu=5$, respectively.  The results of the block-level treatment assignment are visualized in Figure \ref{fig:RE_andnoRE_comparison}

\begin{table} [h!]
    \caption{Comparison of ordinary least squares (OLS), difference-in-differences (DID) and spatiotemporal double ML (DML) for {\bf pixel-level treatment allocation and smoothness $\nu=1$}.  DML methods vary by excluding spatial information (X), using spatial coordinates (XS) and spatial basis functions (XSZ) in BART fits and performing DML with and without cross-fitting (CF).  Methods for estimating the treatment effect are compared using bias, mean squared error (MSE), confidence interval (CI) length and coverage of 95\% intervals. } 
    \renewcommand{\arraystretch}{1}
    \begin{center}\begin{tabular}{lcccc}
    \toprule
        \textbf{Method} & \textbf{Bias} & \textbf{MSE} & \textbf{CI length} & \textbf{Coverage}  \\
    \midrule
        OLS & $0.812$ & $1.090$ & $1.571$ & $0.483$ \\
        DID  & $0.570$ & $0.541$ & $1.826$ & $0.775$  \\
        DML no spatial no cross & $0.562$ & $0.522$ & $1.739$ & $0.775$ \\
        DML no spatial with cross & $0.695$ & $0.700$ & $1.812$ & $0.700$ \\
        DML with spatial no cross & $-0.328$ & $0.182$ & $0.948$ & $0.650$ \\
        DML with spatial with cross & $0.344$ & $0.213$ &$1.261$ & $0.825$ \\
        Wendland without cross & $-0.494$ & $0.294$ & $0.790$ & $0.267$ \\
        Wendland with cross & $0.187$ & $0.091$ & $1.055$ & $0.917$ 
    \end{tabular} \end{center}
    \label{tab:noRE-smooth1}
\end{table}

\begin{table} [h!]
    \caption{Comparison of ordinary least squares (OLS), difference-in-differences (DID) and spatiotemporal double ML (DML) for {\bf pixel-level treatment allocation and smoothness $\nu=5$}.  DML methods vary by excluding spatial information (X), using spatial coordinates (XS) and spatial basis functions (XSZ) in BART fits and performing DML with and without cross-fitting (CF).  Methods for estimating the treatment effect are compared using bias, mean squared error (MSE), confidence interval (CI) length and coverage of 95\% intervals. .} 
    \renewcommand{\arraystretch}{1}
    \begin{center}\begin{tabular}{lcccc}
    \toprule
        \textbf{Method} & \textbf{Bias} & \textbf{MSE} & \textbf{CI length} & \textbf{Coverage}  \\
    \midrule
        OLS & $0.835$ & $1.308$ & $1.573$ & $0.475$ \\
        DID  & $0.541$ & $0.589$ & $1.832$ & $0.750$  \\
        DML no spatial no cross & $0.435$ & $0.434$ & $1.619$ & $0.767$ \\
        DML no spatial with cross & $0.615$ & $0.625$ & $1.714$ & $0.675$ \\
        DML with spatial no cross & $-0.834$ & $0.744$ & $0.565$ & $0.033$ \\
        DML with spatial with cross & $0.140$ & $0.081$ &$0.900$ & $0.892$ \\
        Wendland without cross & $-0.934$ & $0.908$ & $0.472$ & $0.000$ \\
        Wendland with cross & $0.017$ & $0.037$ & $0.709$ & $0.967$ 
    \end{tabular} \end{center}
    \label{tab:noRE-smooth5}
\end{table}

\end{document}